\documentclass[aps,prl,reprint,superscriptaddress]{revtex4-1}

\usepackage{graphicx}
\usepackage{dcolumn}
\usepackage{bm}
\usepackage{mathrsfs}
\usepackage[T1]{fontenc}
\usepackage[utf8]{inputenx}
\usepackage{amsmath}
\usepackage{verbatim}
\usepackage{graphicx}
\usepackage{xcolor}
\usepackage{subfigure}
\usepackage{booktabs} 
\usepackage{amssymb}
\usepackage{mathtools}
\usepackage{amsthm}

\begin{document}


\title{Response Theory via Generative Score Modeling}

\author{Ludovico Theo Giorgini}
\thanks{All the authors contributed equally}
\affiliation{Nordita, Royal Institute of Technology and Stockholm University, Stockholm 106 91, Sweden}
\email{ludovico.giorgini@su.se}

\author{Katherine Deck}
\thanks{All the authors contributed equally}
\affiliation{Climate Modeling Alliance, California Institute of Technology}

\author{Tobias Bischoff}
\thanks{All the authors contributed equally}
\affiliation{8362 Solutions, Pasadena, CA, USA}

\author{Andre Souza}
\thanks{All the authors contributed equally}
\affiliation{Massachusetts Institute of Technology, Cambridge, Massachusetts, United States}

\date{\today}

\begin{abstract}
We introduce an approach for analyzing the responses of dynamical systems to external perturbations that combines score-based generative modeling with the Generalized Fluctuation-Dissipation Theorem (GFDT). The methodology enables accurate estimation of system responses, including those with non-Gaussian statistics. We numerically validate our approach using time-series data from three different stochastic partial differential equations of increasing complexity: an Ornstein-Uhlenbeck process with spatially correlated noise, a modified stochastic Allen-Cahn equation, and the 2D Navier-Stokes equations. We demonstrate the improved accuracy of the methodology over conventional methods and discuss its potential as a versatile tool for predicting the statistical behavior of complex dynamical systems.
\end{abstract}

\maketitle

The study of high-dimensional dynamical systems is essential for advancing our understanding of various complex phenomena \citep[e.g.,][]{Mucha2010, Halu2013, Jumper2021, Brunton2019,barbier2020high, singh2021efficient, Dorogovtsev2008, Barabasi1999}. These systems are characterized by numerous interacting degrees of freedom, manifesting feedback mechanisms across spatial and temporal scales. Notable examples of such complexity are observed in climate modeling \citep[e.g.,][]{ghil2020physics}, where feedback mechanisms lead to self-sustained spatio-temporal patterns, such as the El Niño Southern Oscillation (ENSO) in the tropical Pacific Ocean, the Asian Monsoon (particularly prominent in South and East Asia), the Indian Ocean Dipole, and the Madden-Julian Oscillation in the Indian and Pacific Oceans \citep{timmermann2018nino, martin2021influence}, among others. Similar complexities and the need to understand feedback mechanisms can be observed in other areas, such as financial markets, \citep[e.g.,][]{Badwan2022} or neuroscience \citep[e.g.,][]{Hunt2021, Sporns2021}.

A central challenge is to characterize the causal relationships among these degrees of freedom without prior knowledge of the underlying evolution laws \citep{bozorgmagham2015causality,lagemann2023deep,wismuller2021large,keyes2023stochastic}. Causal inference seeks to unambiguously determine whether the behavior of one variable has been influenced by another based on observed time series data. This estimation process is challenging due to the complexity and high dimensionality of these systems \citep{aurell2016causal,friedrich2011approaching}.

In recent decades, numerous methodologies have been developed to infer causal relationships directly from data \citep[e.g.,][]{granger1969investigating,schreiber2000measuring,pearl2009causal,camps2023discovering,kaddour2022causal}, but calculating a system’s response to small perturbations using linear response theory has become a dominant approach \cite{baldovin2022extracting, Cecconi_2023, falasca2023causal}. This method allows for evaluating responses without actually perturbing the system, leveraging instead the analysis of unperturbed dynamics \citep{baldovin2020understanding,lucarini2018revising}. It is underpinned by the generalized Fluctuation-Dissipation Theorem (GFDT), an extension of the classical Fluctuation-Dissipation Theorem (FDT) \cite{marconi2008fluctuation}. A central obstacle in applying linear response theory is obtaining the system’s unperturbed invariant distribution, i.e., the distribution of a system's attractor \citep{daum2005nonlinear,sjoberg2009fokker, ClimateSensitivityviaaNonparametricFluctuationDissipationTheorem}.

The invariant distribution is often approximated by assuming a multivariate Gaussian distribution ansatz to circumvent this issue \citep{ClimateResponseandFluctuationDissipation}. This approximation is reasonable when examining coarse-grained (spatially or temporally) variables or particular physical observables \citep{ReconcilingNonGaussianClimateStatisticswithLinearDynamics, GRITSUN201762}. The limits of this approximation are evident in atmospheric science, where observational data from  wind and rainfall exhibit intermittency and other non-Gaussian features \citep{proistosescu2016identification,loikith2015short, branicki2012quantifying,  WhyDoPrecipitationIntensitiesTendtoFollowGammaDistributions}. Another context for non-Gaussian statistics comes from neuroscience, where the brain’s response to stimuli and the resulting patterns of neural activity display non-Gaussian properties \citep{kriegeskorte2021neural}.

Our work leverages recent advancements in score-based generative modeling \citep[e.g.,][and follow-on studies]{Vincent2011, Ho2020, Song2019, song2020score, Bischoff2023, liu2023genphys} to accurately represent a high-dimensional, nonlinear dynamical system’s invariant distribution via a “score function,” the gradient of the logarithm of a system’s invariant distribution. Score models are used primarily as generative models, i.e., for sampling from the data distribution. Still, they have other applications, such as estimating the dimensionality of the data distribution, see \citep{stanczuk2023diffusion}. This study demonstrates another use case beyond sample generation and in an area where machine-learning approaches have not yet been fully utilized: a trained score model can be used to compute response functions using data from dynamical systems.

Using trajectories of the dynamical system, we train a model that approximates the score of the steady-state, or invariant, distribution. The model is a neural network based on spatial convolutions that can capture high-dimensional data distribution statistics (e.g., \citep{Ronneberger2015, adcock2020deep, beneventano2021deep}). Like the Gaussian approximation approach, this method only requires time-series data. In this work, we apply the score-based method to three dynamical systems and compare its performance to other standard approaches for computing response functions. This approach outperforms the traditional Gaussian approximation when the underlying system is nonlinear.  

\newpage
\textit{Problem description. --} We focus on systems described by an evolution equation
\begin{equation}
    \partial_t u= F(u) + \xi,
    \label{eq:motion}
\end{equation}
where $u$ is the system state, which varies as a function of location in space and time and belongs to a space $\mathcal{X}$. The function $F: \mathcal{X} \rightarrow \mathcal{X}$ is a map, and $\xi$ is a noise term, which can be taken to be zero to yield a deterministic system. However, we emphasize that the only requirements for applying the methodology here are time-series data, and one does not need to know an underlying evolution law such as Equation~(\ref{eq:motion}), which represents systems of both ordinary or partial differential equations. 

Introducing a finite-dimensional analog facilitates the connection with computations. When computations are performed on a grid,  we use $u_i(t) \equiv u(x_i, t)$ as shorthand to denote the values of $u$ at grid location $x_i$ at time $t$, where $i \in [1,N^2]$, and $N^2$ is the number of grid points. Thinking of $u_i$ as the components of a vector in $\mathbb{R}^{N^2}$, we also use $\vec{u}(t)$ to denote the vector of values $u$ at time $t$ and each grid location $x_i$. We use similar notation for other functions as well, e.g. if $\vec{f}: \mathbb{R}^{N^2} \rightarrow \mathbb{R}^{N^2}$, then $f_i(u(t))$ denotes the value of $\vec{f}(\vec{u})$ at grid location $x_i$ and time $t$.  

We consider statistically stationary (steady-state) systems with a smooth invariant probability density function $\rho = \rho(u_1, u_2, ..., u_N^2)$. We will use the GFDT, which is defined in Equation~\eqref{eq:GFDT}, to understand how a small initial perturbation applied at time $t$ alters the expected value of a system's state at a later time $t+\mathcal{T}$, compared to its unperturbed expected state.  This relative change in the state after a time $\mathcal{T}$, $\delta \vec{u}(\mathcal{T})$, has the expectation
\begin{equation}
    \langle \delta \vec{u}(\mathcal{T}) \rangle_p \equiv \langle \vec{u}(\mathcal{T}) \rangle_{\mathcal{J}'} - \langle \vec{u} \rangle_{\rho},
    \label{eq:deltax}
\end{equation}
where $\langle \cdot \rangle_\rho$ denotes an expectation over the steady-state attractor distribution $\rho$, $\langle \cdot \rangle_{\mathcal{J}'}$ denotes an expectation over the joint density $\mathcal{J}'(\vec{u}(t), \vec{v}(t+\mathcal{T}) \lvert \vec{v}(t)  = \vec{u}(t) + \delta \vec{u}(t))$, for a fixed perturbation $\delta \vec{u}(t)$ at time $t$. The ensemble average $\langle \cdot \rangle_p$ is defined by the right-hand side of Equation~\eqref{eq:deltax}. Below, we will use $\langle \cdot \rangle_\mathcal{J}$ to denote an expectation over the joint density $\mathcal{J}(\vec{u}(t), \vec{v}(t+\mathcal{T}) \lvert \vec{v}(t)  = \vec{u}(t)),$ which can be written as $\mathcal{J}(\vec{u}(t), \vec{u}(t+\mathcal{T}))$.

For a small perturbation to $u$ at grid location $j$, denoted by $\delta u_j$, we wish to quantify the change in the average of $u_i$ at a later time $\mathcal{T}$ (the discrete analog of Equation~\eqref{eq:deltax}). The GFDT states that this can be evaluated using a matrix-valued quantity called the response function $R_{ij}$ (we also refer to it as the response matrix) \cite{marconi2008fluctuation} as
\begin{equation}
     R_{ij}(\mathcal{T}) \equiv \frac{\langle \delta u_i (\mathcal{T}) \rangle_p}{ \delta u_j(0)} =-\left \langle u_i(\mathcal{T}) s_j(\vec{u}(0))\right \rangle_{\mathcal{J}},
     \label{eq:GFDT}
 \end{equation}
where the score function of the steady-state distribution has the usual definition
\begin{equation}
    \vec{s} = \nabla \, \ln \rho.
    \label{eq:score}
\end{equation}
Since we assume that we have access to time-series data, the problem of estimating the response function is thus reduced to estimating the score function $\vec{s}(\vec{u})$ of the steady-state distribution $\rho$. In deterministic dynamical systems, this invariant distribution is typically singular almost everywhere on the attractor, and it is necessary to introduce Gaussian noise into the system to make $\rho$ sufficiently smooth to apply the GFDT \cite{gritsun2007climate} effectively.

If the steady-state distribution is a multivariate Gaussian distribution, the response matrix is given by
\begin{equation}
    \bm{R}(\mathcal{T}) = \bm{C}(\mathcal{T})\bm{C}^{-1}(0),
    \label{eq:response_linear}
\end{equation}
where $\bm{C}(\mathcal{T})$ is the correlation matrix, with elements $C_{ij}(\mathcal{T})=\langle u_i(\mathcal{T}) u_j(0) \rangle_{\mathcal{J}}$. The property that $R_{ij}(0) = \delta_{ij}$, where $\delta_{ij}$ is the Kronecker delta, is a general property that holds for the discrete response matrix, see the Supplementary Information (SI). Constructing the correlation matrix is straightforward, allowing Equation
(\ref{eq:response_linear}) to serve as a convenient method for estimating the response matrix when the underlying steady-state distribution is well-approximated by a Gaussian distribution.

The response function can also be estimated using numerical simulations of the dynamical system. This estimation involves taking an ensemble of initial conditions, perturbing them, evolving the unperturbed and perturbed system forward in time, and then taking an ensemble average. However, this must be repeated for each perturbation of interest, which involves repeated potentially expensive simulations. It also requires knowledge of the equations of motion.

Here, we employ score-based generative modeling to derive the response function directly from data. The approach uses denoising score matching to facilitate the parameterization of the dynamical system's attractor score function directly, following methodologies suggested by Song et al. \citep{Song2019, song2020score}. Specifically, we use a U-Net to represent the score-function \cite{Ronneberger2015, Bischoff2023}, but other choices of architecture are necessary depending on the application. Our approach only requires data and does not require approximating the steady-state distribution as Gaussian or additional numerical simulations for different perturbations. Refer to the SI for additional details on the architecture, training the model, obtaining the response function, and practical considerations. 

\begin{figure*}[htbp]
\centering
\centerline{\includegraphics[width=\linewidth]{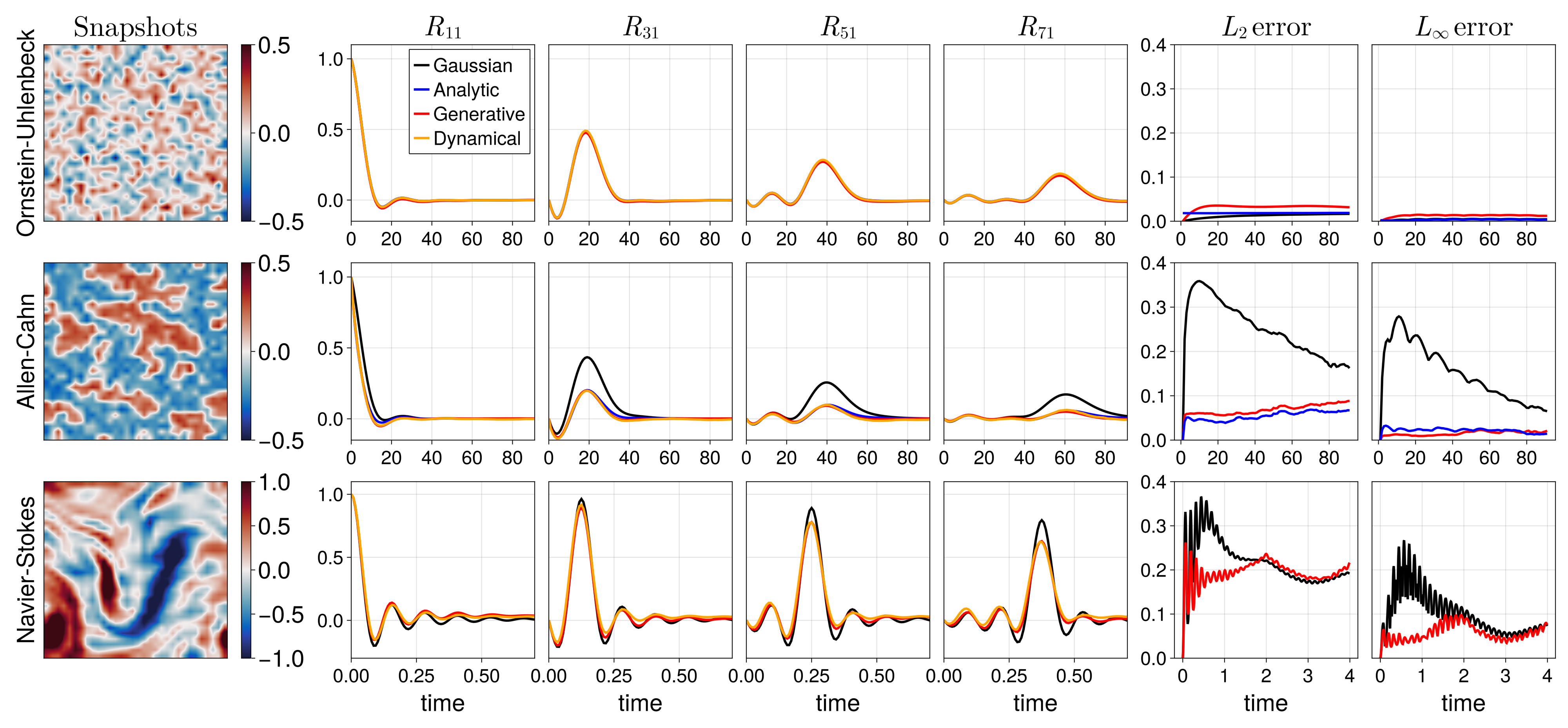}}
\caption{\textbf{Column 1: Example snapshots of the Three Systems}. In the \textbf{top row}, Ornstein-Uhlenbeck PDE, in the \textbf{middle row}  Allen-Cahn, in the \textbf{bottom row} Navier-Stokes. \textbf{Columns 2-5: Response functions for Three Systems}. The response to a perturbation at the pixel component $i=1$ is evaluated at various pixel components with coordinates indicated at each panel. All evaluated components align with the advection direction. Responses are computed by integrating the dynamical system (orange lines, ``Dynamical'') via Gaussian approximation (black lines, ``Gaussian''), using the score function derived from generative modeling (red lines, ``Generative''), and using the analytic score function (blue lines, ``Analytic''). \textbf{Columns 6-7: Response Function Errors for the Three Systems.} Here, we show both the RMS error and the Infinity Norm error of the Response function as a function of time. Here we use the Dynamic Response function as the ``ground truth''. We see that the Generative response (red) generally outperforms the Gaussian response (blue) for the nonlinear cases (Allen-Cahn and Navier-Stokes), especially in the Infinity Norm, and performs similarly for the linear system (Ornstein-Uhlenbeck). The response function computed with the exact score does not have zero error relative to the dynamical response function because of errors introduced in numerical discretization and empirical averaging.}.
\label{figure}
\end{figure*}

We apply the proposed method to three different dynamical systems. We compare the response function computed in this way (Generative) to a response computed by solving the perturbed dynamical system (Dynamical), a response using an analytically derived score function (Analytical) for the systems for which it is possible to derive one, and the response obtained through the Gaussian approximation (Gaussian). 

Before describing the three systems under study, we note some commonalities. Each represents a stochastic partial differential equation for a scalar variable. We take the spatial domain to be two-dimensional and periodic in each direction. The state field $u$ maps a two-dimensional location into a scalar field, which we represent with a spatial discretization of $N^2$ degrees of freedom (or pixels, when representing the field as an image). In our study, we used $N = 32$ pixels. We use a coarse resolution so that comparing to the Gaussian approximation is directly tractable as it requires inverting a $32^2 \times 32^2$ matrix. An advection term, $ A \partial_x u $, where $A$ is a constant, was added because it introduces a preferred direction of information propagation and, therefore, a direction of causality. The systems differ in their complexity: the first system has Gaussian statistics, the second system is non-Gaussian but has analytically tractable score function, and the last system is non-Gaussian, does not have an analytically tractable score function, and has ``hidden dynamics'' in the forcing.

\textit{An Ornstein-Uhlenbeck Stochastic PDE. --} Our first system is an Ornstein-Uhlenbeck (OU) system \cite{uhlenbeck1930theory} with spatially correlated noise, \cite{maller2009ornstein, griffin2006inference, frank2000multivariate, vatiwutipong2019alternative}. The steady-state trajectories of the system exhibit Gaussian statistics, which makes the Gaussian approximation exact, and hence serves as a baseline for testing our method. 

Here, we consider a multidimensional version with an advection term and spatially correlated noise:
\begin{equation}\label{eq:ou_eom}
    \partial_t u = - \lambda \Sigma   u + \kappa \Delta u + A \partial_x u + \epsilon \Sigma^{1/2} \xi,
\end{equation}
where $ u = u(x,y,t) $ is the unknown field which depends on space $ x, y $, and time $ t $, $ \Sigma \equiv (1 - \Delta)^{-1} $ is an inverse Helmholtz operator, $ A = 2 \times 10^{-2}, \epsilon = 2 \times 10^{-3/2}, \lambda = 4 \times 10^{-2}$ are constant scalars, and $ \xi $ represents space-time noise with covariance
\begin{equation}
    \mathbb{C}\text{ov}\left[\xi(x,y,t)\right] = \delta(x - x', y - y', t - t'),
\end{equation}
where $ \delta(x,y,t) $ is the Dirac delta function. In this case, an analytically tractable score function is available
\begin{equation}
    \label{eq:ou_analytic_score}
    s(u) = \frac{2}{\epsilon^2} \left( - \lambda u + \kappa \Delta u - \kappa \Delta^2 u \right), 
\end{equation}
which doesn't depend on the advective term $A \partial_x u$.

\textit{A modified Allen-Cahn system. --} Our second system is the Allen-Cahn system, an example of a reaction-diffusion equation used for modeling phase separation in multi-component systems \citep[see][who studied applications to binary alloys]{allen1972ground}. Reaction-diffusion equations \citep{kolmogorov1study} are partial differential equations combining a diffusive term with a (non)linear reaction term. They are commonly used to study pattern formation (e.g., \citep{turing1990chemical, callahan1999pattern, kondo2010reaction}) and chemical processes. 

We considered a modified version of the Allen-Cahn system to include an advection term and spatially correlated noise, as follows,
\begin{equation}
\partial_t u = \kappa \Delta u - A\partial_x u + \alpha \Sigma \left( u (1 - u^2) \right) + \epsilon \Sigma^{1/2}  \xi,
\label{eq:allencahn_eom}
\end{equation}
where $\Sigma, A, \epsilon, u, \xi$ are same as those used in the OU system.  We choose parameters $\alpha = 16$, $\kappa= 2.5 \times 10^{-4}$. The system has an analytically tractable score function, 
\begin{equation}
\label{eq:allencahn_analytic_score}
s(u) =  \frac{2}{\epsilon^2} \left[\alpha u(1-u^2) + \kappa (\Delta u - \Delta^2 u) \right],
\end{equation}
which, similar to the OU example, doesn't depend on the advective term $A \partial_x u$. The steady-state trajectories of this system exhibit bimodal statistics, and hence provides an example where the Gaussian approximation to the response function is expected to fail. 

For both the OU and Allen-Cahn systems, the analytic score function and the machine-learned score will differ by a constant factor due to the difference between the Kronecker delta and the Dirac delta functions. In both examples, the analytic score function is $32^2$ larger. We divide the analytic score by this factor for comparison with the machine-learned score. Refer to the SI for the derivation of the analytic score functions.

\textit{Navier-Stokes equation. --} We next study a version of the 2D Navier-Stokes model as described in \cite{Flierl_Souza_2024}, where we take the state variable to be the vorticity, $u$. This system was chosen as a stepping-stone towards the more complex fluids observed in climate systems; however, the vorticity statistics are not strongly non-Gaussian due to the chosen resolution as would be expected from a coarse-grained system, \cite{ReconcilingNonGaussianClimateStatisticswithLinearDynamics}. Hence, we expect good performance from the Gaussian approximation to the response function. The data-generating equation is
\begin{align}\label{eq:navierstokes_eom}
\partial_t u  &= \partial_x \psi \partial_y u - \partial_x u \partial_y \psi - A \partial_x u +  \mathcal{D} u  + \varsigma
\\
\psi &= \Delta^{-1} u
\\
\mathcal{D}u &= - \nu_h \Delta^{-2}u -  \nu \Delta^2u - \nu_0 \int u \text{d} x \text{d} y 
\end{align}
where $\mathcal{D}$ is a dissipation operator with constants ($\nu_h = 10^{-2}$, $\nu = 10^{-5}$, $\nu_0 = (2\pi)^{-2}$), $A = \pi$ is a mean advection term, and $\varsigma$ is a random wave forcing from \cite{Flierl_Souza_2024}. The random wave forcing introduces hidden dynamics into the system, thus the score function is estimated using only partial state information. The inverse Laplacian is defined to preserve the mean of the original variable. 

\textit{Results. --}
We first generate time series data corresponding to Equations~\eqref{eq:ou_eom}, ~\eqref{eq:allencahn_eom}, and Equation~\eqref{eq:navierstokes_eom}. These data are used to compute response functions and the training sets for the score model. To create the training data, we determine autocorrelation times for each and subsample the trajectories to create statistically independent snapshots. Figure~(\ref{figure}) Column (1) shows training data samples of the numerically simulated systems.
All details regarding the numerical simulations (time-stepping, spatial discretization), the datasets and the training of the score models are provided in the SI, along with an in-depth comparison between data samples generated via numerical simulation of the Allen-Cahn system and those produced using the learned score function of this system through reverse diffusion. Training the neural network took a few hours to a few days to for each score model on a single Nvidia A100. 

Owing to the periodic boundary conditions and symmetries of the systems under study, the response matrix elements exhibit translational invariance in the $x$ and $y$-directions. However, due to the advection term, the response function in the $x$-direction depends on both the distance and the direction (sign) between pixels while the $y$-direction only depends on the distance.
This invariance implies that the index $i$ in the response matrix $R_{ij}$ can be arbitrarily fixed. We chose $j=1$ and calculated the response matrix elements along the direction of advection (denoted by $i$), as this is the direction information propagation, where we observe a more significant response to perturbations at pixel component $i = 1$. The resulting responses of nearby pixels as a function of time for the three systems are presented in Figure~(\ref{figure}) Columns (2-5). The generative model is consistent with the dynamical response function (``our ground truth'') in all cases. For the OU system, the Gaussian response function also matches the ground truth; this is expected as this system exhibits Gaussian statistics.  For the Allen-Cahn system, it is clear that the Gaussian approximation is a poor approximation to the response function. And lastly, for the Navier-Stokes system we see improvement in using the generative score over the Gaussian response, despite the coarse Navier-Stokes system being quasi-Gaussian.

In Figure~(\ref{figure}) Columns (6-7), we compared the error over the entire state field between the Generative, Gaussian, and Analytic (for the systems where it is available) response functions with the Dynamical response function considered as ``ground truth'' for the three systems as a function of time. We used two different metrics to determine the errors: the $L^2$ and the $L^\infty$ norm.
The $L^2$ norm measures the average error across all pixels in the domain, providing a sense of the overall error distribution. It is helpful in understanding the general accuracy of the response function over the entire field. On the other hand, the $L^\infty$ norm measures the maximum error at any single pixel, highlighting the worst-case error. This norm is particularly relevant to ensure the minimization of the largest deviations.
From the comparison in Figure~(\ref{figure}), it is evident that the Generative response function generally outperforms the Gaussian response in the nonlinear cases (Allen-Cahn and Navier-Stokes), especially in terms of the $L^\infty$ norm at early to intermediate times. This observation indicates that the Generative model is better at capturing the extreme responses at individual pixels. The Gaussian approximation is exact for the Ornstein-Uhlenbeck system, which has a Gaussian steady-state distribution, and both the Generative and Gaussian response functions perform similarly.

\textit{Drawbacks and Practical Considerations. --} While score-based generative models offer a powerful tool for estimating response functions, there are practical challenges. One  issue is that training these models requires substantial data or a sufficiently good inductive bias in the parameterized model for the score function. Additionally, setting up and training neural networks can often require significant resources, engineering, and time. Expertise in training neural networks is essential. For example, suppose the data or training time is insufficient, and one is using a U-Net to represent the score function. In that case, the generative response function tends to match the Gaussian approximation's performance due to the U-Net's linear bypass connections.  A further drawback with any machine-learning method is the lack of performance ``out-of-sample''; indeed, in the SI, we demonstrate that the score model for the Allen-Cahn system performs poorly when evaluated on data dissimilar from the training data.

For many applications, especially those mentioned in the introduction, several additional steps may need to be taken to apply the methods from this study effectively. One crucial step is the incorporation of domain-specific knowledge into the design of the neural network architecture. This will significantly improve the model's inductive bias and alleviate the data burden. Additionally, domain-specific challenges such as non-stationarity or incomplete state information must be incorporated \citep{LowFrequencyClimateResponseandFluctuationDissipationTheoremsTheoryandPractice}. This is particularly salient for climate data, which exhibits diurnal, seasonal, and longer timescale variations \citep{GERSHGORIN20101741, Majda2019}. As always, it is crucial to validate the trained model with known benchmarks or analytical solutions.

\textit{Conclusions. --} In this study, we use score-based diffusion models to compute response functions, offering a purely data-driven method to calculate accurate response functions. We validated this approach on a system with Gaussian statistics (Ornstein-Uhlenbeck) and two dynamical systems with more complex distributions: a reaction-diffusion equation (Allen-Cahn) and the Navier-Stokes equations. The former is a class of SPDEs that have applications across biology, chemistry, and physics, and the latter is a system exhibiting some of the characteristics of more complex fluids like the Earth's atmosphere or ocean. We compared the response function of these systems computed in multiple ways: using an \textit{analytic} expression for the score, using the \textit{generative} score function, using a \textit{dynamical} approach by introducing perturbations to initial conditions and solving the equations of motion, and using a \textit{Gaussian} approximation. For the nonlinear cases, we found that the Gaussian response function over-predicted the response to neighboring perturbations and that the generative response function agreed well with both the dynamical and analytical response functions. Compared with the Gaussian response, the RMSE of the generative response function had significantly reduced error. This suggests that score-based models are a useful tool for studying system response functions, especially when the dynamics are strongly nonlinear.

The code utilized in this work is publicly available for further study and replication at: \url{https://anonymous.4open.science/r/PrivateGenLRT-052E/generate/allen_cahn_model.jl}. We used the publicly available codebase \url{https://github.com/CliMA/CliMAgen.jl/} as the backbone for training the models.

\vspace{\baselineskip}
\begin{acknowledgments}
The authors would like to thank three anonymous reviewers for their suggestions which greatly enhanced the scope, quality, and presentation of the letter. LTG gratefully acknowledges support from the Swedish Research Council (Vetenskapsradet) Grant No. 638-2013-9243. KD acknowledges support from Schmidt Sciences and the Cisco Foundation. AS and LTG acknowledge support from Schmidt Sciences through the Bringing Computation to the Climate Challenge, an MIT Climate Grand Challenge Project. AS thanks Fabrizio Falasca and Glenn Flierl for their invaluable suggestions in improving preliminary versions of this work. TB acknowledges the support of the community at Livery Studio and useful discussions with Bryan Riel on generative modeling.
\end{acknowledgments}

\bibliography{references}

\begin{thebibliography}{67}%
\makeatletter
\providecommand \@ifxundefined [1]{%
 \@ifx{#1\undefined}
}%
\providecommand \@ifnum [1]{%
 \ifnum #1\expandafter \@firstoftwo
 \else \expandafter \@secondoftwo
 \fi
}%
\providecommand \@ifx [1]{%
 \ifx #1\expandafter \@firstoftwo
 \else \expandafter \@secondoftwo
 \fi
}%
\providecommand \natexlab [1]{#1}%
\providecommand \enquote  [1]{``#1''}%
\providecommand \bibnamefont  [1]{#1}%
\providecommand \bibfnamefont [1]{#1}%
\providecommand \citenamefont [1]{#1}%
\providecommand \href@noop [0]{\@secondoftwo}%
\providecommand \href [0]{\begingroup \@sanitize@url \@href}%
\providecommand \@href[1]{\@@startlink{#1}\@@href}%
\providecommand \@@href[1]{\endgroup#1\@@endlink}%
\providecommand \@sanitize@url [0]{\catcode `\\12\catcode `\$12\catcode `\&12\catcode `\#12\catcode `\^12\catcode `\_12\catcode `\%12\relax}%
\providecommand \@@startlink[1]{}%
\providecommand \@@endlink[0]{}%
\providecommand \url  [0]{\begingroup\@sanitize@url \@url }%
\providecommand \@url [1]{\endgroup\@href {#1}{\urlprefix }}%
\providecommand \urlprefix  [0]{URL }%
\providecommand \Eprint [0]{\href }%
\providecommand \doibase [0]{http://dx.doi.org/}%
\providecommand \selectlanguage [0]{\@gobble}%
\providecommand \bibinfo  [0]{\@secondoftwo}%
\providecommand \bibfield  [0]{\@secondoftwo}%
\providecommand \translation [1]{[#1]}%
\providecommand \BibitemOpen [0]{}%
\providecommand \bibitemStop [0]{}%
\providecommand \bibitemNoStop [0]{.\EOS\space}%
\providecommand \EOS [0]{\spacefactor3000\relax}%
\providecommand \BibitemShut  [1]{\csname bibitem#1\endcsname}%
\let\auto@bib@innerbib\@empty
\bibitem [{\citenamefont {Mucha}\ \emph {et~al.}(2010)\citenamefont {Mucha}, \citenamefont {Richardson}, \citenamefont {Macon}, \citenamefont {Porter},\ and\ \citenamefont {Onnela}}]{Mucha2010}%
  \BibitemOpen
  \bibfield  {author} {\bibinfo {author} {\bibfnamefont {P.~J.}\ \bibnamefont {Mucha}}, \bibinfo {author} {\bibfnamefont {T.}~\bibnamefont {Richardson}}, \bibinfo {author} {\bibfnamefont {K.}~\bibnamefont {Macon}}, \bibinfo {author} {\bibfnamefont {M.~A.}\ \bibnamefont {Porter}}, \ and\ \bibinfo {author} {\bibfnamefont {J.-P.}\ \bibnamefont {Onnela}},\ }\href@noop {} {\bibfield  {journal} {\bibinfo  {journal} {Science}\ }\textbf {\bibinfo {volume} {328}},\ \bibinfo {pages} {876} (\bibinfo {year} {2010})}\BibitemShut {NoStop}%
\bibitem [{\citenamefont {Halu}\ \emph {et~al.}(2013)\citenamefont {Halu}, \citenamefont {Mondragón}, \citenamefont {Panzarasa},\ and\ \citenamefont {Bianconi}}]{Halu2013}%
  \BibitemOpen
  \bibfield  {author} {\bibinfo {author} {\bibfnamefont {A.}~\bibnamefont {Halu}}, \bibinfo {author} {\bibfnamefont {R.~J.}\ \bibnamefont {Mondragón}}, \bibinfo {author} {\bibfnamefont {P.}~\bibnamefont {Panzarasa}}, \ and\ \bibinfo {author} {\bibfnamefont {G.}~\bibnamefont {Bianconi}},\ }\href@noop {} {\bibfield  {journal} {\bibinfo  {journal} {PLoS ONE}\ }\textbf {\bibinfo {volume} {8}},\ \bibinfo {pages} {e78293} (\bibinfo {year} {2013})}\BibitemShut {NoStop}%
\bibitem [{\citenamefont {Jumper}\ \emph {et~al.}(2021)\citenamefont {Jumper} \emph {et~al.}}]{Jumper2021}%
  \BibitemOpen
  \bibfield  {author} {\bibinfo {author} {\bibfnamefont {J.}~\bibnamefont {Jumper}} \emph {et~al.},\ }\href@noop {} {\bibfield  {journal} {\bibinfo  {journal} {Nature}\ }\textbf {\bibinfo {volume} {596}},\ \bibinfo {pages} {583} (\bibinfo {year} {2021})}\BibitemShut {NoStop}%
\bibitem [{\citenamefont {Brunton}\ \emph {et~al.}(2019)\citenamefont {Brunton}, \citenamefont {Noack},\ and\ \citenamefont {Koumoutsakos}}]{Brunton2019}%
  \BibitemOpen
  \bibfield  {author} {\bibinfo {author} {\bibfnamefont {S.~L.}\ \bibnamefont {Brunton}}, \bibinfo {author} {\bibfnamefont {B.~R.}\ \bibnamefont {Noack}}, \ and\ \bibinfo {author} {\bibfnamefont {P.}~\bibnamefont {Koumoutsakos}},\ }\href@noop {} {\bibfield  {journal} {\bibinfo  {journal} {Annual Review of Fluid Mechanics}\ }\textbf {\bibinfo {volume} {52}},\ \bibinfo {pages} {477} (\bibinfo {year} {2019})}\BibitemShut {NoStop}%
\bibitem [{\citenamefont {Barbier}(2020)}]{barbier2020high}%
  \BibitemOpen
  \bibfield  {author} {\bibinfo {author} {\bibfnamefont {J.}~\bibnamefont {Barbier}},\ }\href@noop {} {\bibfield  {journal} {\bibinfo  {journal} {arXiv preprint arXiv:2010.14863}\ } (\bibinfo {year} {2020})}\BibitemShut {NoStop}%
\bibitem [{\citenamefont {Singh}\ \emph {et~al.}(2021)\citenamefont {Singh}, \citenamefont {Wang}, \citenamefont {Cole},\ and\ \citenamefont {Ching}}]{singh2021efficient}%
  \BibitemOpen
  \bibfield  {author} {\bibinfo {author} {\bibfnamefont {M.~F.}\ \bibnamefont {Singh}}, \bibinfo {author} {\bibfnamefont {C.}~\bibnamefont {Wang}}, \bibinfo {author} {\bibfnamefont {M.~W.}\ \bibnamefont {Cole}}, \ and\ \bibinfo {author} {\bibfnamefont {S.}~\bibnamefont {Ching}},\ }\href@noop {} {\bibfield  {journal} {\bibinfo  {journal} {arXiv preprint arXiv:2104.02827}\ } (\bibinfo {year} {2021})}\BibitemShut {NoStop}%
\bibitem [{\citenamefont {Dorogovtsev}\ \emph {et~al.}(2008)\citenamefont {Dorogovtsev}, \citenamefont {Goltsev},\ and\ \citenamefont {Mendes}}]{Dorogovtsev2008}%
  \BibitemOpen
  \bibfield  {author} {\bibinfo {author} {\bibfnamefont {S.~N.}\ \bibnamefont {Dorogovtsev}}, \bibinfo {author} {\bibfnamefont {A.~V.}\ \bibnamefont {Goltsev}}, \ and\ \bibinfo {author} {\bibfnamefont {J.~F.~F.}\ \bibnamefont {Mendes}},\ }\href@noop {} {\bibfield  {journal} {\bibinfo  {journal} {Reviews of Modern Physics}\ }\textbf {\bibinfo {volume} {80}},\ \bibinfo {pages} {1275} (\bibinfo {year} {2008})}\BibitemShut {NoStop}%
\bibitem [{\citenamefont {Barabási}\ and\ \citenamefont {Albert}(1999)}]{Barabasi1999}%
  \BibitemOpen
  \bibfield  {author} {\bibinfo {author} {\bibfnamefont {A.-L.}\ \bibnamefont {Barabási}}\ and\ \bibinfo {author} {\bibfnamefont {R.}~\bibnamefont {Albert}},\ }\href@noop {} {\bibfield  {journal} {\bibinfo  {journal} {Science}\ }\textbf {\bibinfo {volume} {286}},\ \bibinfo {pages} {509} (\bibinfo {year} {1999})}\BibitemShut {NoStop}%
\bibitem [{\citenamefont {Ghil}\ and\ \citenamefont {Lucarini}(2020)}]{ghil2020physics}%
  \BibitemOpen
  \bibfield  {author} {\bibinfo {author} {\bibfnamefont {M.}~\bibnamefont {Ghil}}\ and\ \bibinfo {author} {\bibfnamefont {V.}~\bibnamefont {Lucarini}},\ }\href@noop {} {\bibfield  {journal} {\bibinfo  {journal} {Reviews of Modern Physics}\ }\textbf {\bibinfo {volume} {92}},\ \bibinfo {pages} {035002} (\bibinfo {year} {2020})}\BibitemShut {NoStop}%
\bibitem [{\citenamefont {Timmermann}\ \emph {et~al.}(2018)\citenamefont {Timmermann}, \citenamefont {An}, \citenamefont {Kug}, \citenamefont {Jin}, \citenamefont {Cai}, \citenamefont {Capotondi}, \citenamefont {Cobb}, \citenamefont {Lengaigne}, \citenamefont {McPhaden}, \citenamefont {Stuecker} \emph {et~al.}}]{timmermann2018nino}%
  \BibitemOpen
  \bibfield  {author} {\bibinfo {author} {\bibfnamefont {A.}~\bibnamefont {Timmermann}}, \bibinfo {author} {\bibfnamefont {S.-I.}\ \bibnamefont {An}}, \bibinfo {author} {\bibfnamefont {J.-S.}\ \bibnamefont {Kug}}, \bibinfo {author} {\bibfnamefont {F.-F.}\ \bibnamefont {Jin}}, \bibinfo {author} {\bibfnamefont {W.}~\bibnamefont {Cai}}, \bibinfo {author} {\bibfnamefont {A.}~\bibnamefont {Capotondi}}, \bibinfo {author} {\bibfnamefont {K.~M.}\ \bibnamefont {Cobb}}, \bibinfo {author} {\bibfnamefont {M.}~\bibnamefont {Lengaigne}}, \bibinfo {author} {\bibfnamefont {M.~J.}\ \bibnamefont {McPhaden}}, \bibinfo {author} {\bibfnamefont {M.~F.}\ \bibnamefont {Stuecker}},  \emph {et~al.},\ }\href@noop {} {\bibfield  {journal} {\bibinfo  {journal} {Nature}\ }\textbf {\bibinfo {volume} {559}},\ \bibinfo {pages} {535} (\bibinfo {year} {2018})}\BibitemShut {NoStop}%
\bibitem [{\citenamefont {Martin}\ \emph {et~al.}(2021)\citenamefont {Martin}, \citenamefont {Son}, \citenamefont {Butler}, \citenamefont {Hendon}, \citenamefont {Kim}, \citenamefont {Sobel}, \citenamefont {Yoden},\ and\ \citenamefont {Zhang}}]{martin2021influence}%
  \BibitemOpen
  \bibfield  {author} {\bibinfo {author} {\bibfnamefont {Z.}~\bibnamefont {Martin}}, \bibinfo {author} {\bibfnamefont {S.-W.}\ \bibnamefont {Son}}, \bibinfo {author} {\bibfnamefont {A.}~\bibnamefont {Butler}}, \bibinfo {author} {\bibfnamefont {H.}~\bibnamefont {Hendon}}, \bibinfo {author} {\bibfnamefont {H.}~\bibnamefont {Kim}}, \bibinfo {author} {\bibfnamefont {A.}~\bibnamefont {Sobel}}, \bibinfo {author} {\bibfnamefont {S.}~\bibnamefont {Yoden}}, \ and\ \bibinfo {author} {\bibfnamefont {C.}~\bibnamefont {Zhang}},\ }\href@noop {} {\bibfield  {journal} {\bibinfo  {journal} {Nature Reviews Earth \& Environment}\ }\textbf {\bibinfo {volume} {2}},\ \bibinfo {pages} {477} (\bibinfo {year} {2021})}\BibitemShut {NoStop}%
\bibitem [{\citenamefont {Badwan}(2022)}]{Badwan2022}%
  \BibitemOpen
  \bibfield  {author} {\bibinfo {author} {\bibfnamefont {N.}~\bibnamefont {Badwan}},\ }\href@noop {} {\bibfield  {journal} {\bibinfo  {journal} {Macroeconomic Analysis for Economic Growth}\ ,\ \bibinfo {pages} {145}} (\bibinfo {year} {2022})}\BibitemShut {NoStop}%
\bibitem [{\citenamefont {Hunt}\ \emph {et~al.}(2021)\citenamefont {Hunt}, \citenamefont {Tewarie}, \citenamefont {Smith}, \citenamefont {Porcaro}, \citenamefont {Singh},\ and\ \citenamefont {Murphy}}]{Hunt2021}%
  \BibitemOpen
  \bibfield  {author} {\bibinfo {author} {\bibfnamefont {B.~A.}\ \bibnamefont {Hunt}}, \bibinfo {author} {\bibfnamefont {P.~K.}\ \bibnamefont {Tewarie}}, \bibinfo {author} {\bibfnamefont {A.~G.~G.}\ \bibnamefont {Smith}}, \bibinfo {author} {\bibfnamefont {C.}~\bibnamefont {Porcaro}}, \bibinfo {author} {\bibfnamefont {K.~D.}\ \bibnamefont {Singh}}, \ and\ \bibinfo {author} {\bibfnamefont {P.~R.}\ \bibnamefont {Murphy}},\ }\href {\doibase 10.1371/journal.pbio.3001686} {\bibfield  {journal} {\bibinfo  {journal} {PLOS Biology}\ } (\bibinfo {year} {2021}),\ 10.1371/journal.pbio.3001686}\BibitemShut {NoStop}%
\bibitem [{\citenamefont {Sporns}\ and\ \citenamefont {Betzel}(2021)}]{Sporns2021}%
  \BibitemOpen
  \bibfield  {author} {\bibinfo {author} {\bibfnamefont {O.}~\bibnamefont {Sporns}}\ and\ \bibinfo {author} {\bibfnamefont {R.~F.}\ \bibnamefont {Betzel}},\ }\href {\doibase 10.1038/s41593-021-00858-0} {\bibfield  {journal} {\bibinfo  {journal} {Nature Neuroscience}\ } (\bibinfo {year} {2021}),\ 10.1038/s41593-021-00858-0}\BibitemShut {NoStop}%
\bibitem [{\citenamefont {BozorgMagham}\ \emph {et~al.}(2015)\citenamefont {BozorgMagham}, \citenamefont {Motesharrei}, \citenamefont {Penny},\ and\ \citenamefont {Kalnay}}]{bozorgmagham2015causality}%
  \BibitemOpen
  \bibfield  {author} {\bibinfo {author} {\bibfnamefont {A.~E.}\ \bibnamefont {BozorgMagham}}, \bibinfo {author} {\bibfnamefont {S.}~\bibnamefont {Motesharrei}}, \bibinfo {author} {\bibfnamefont {S.~G.}\ \bibnamefont {Penny}}, \ and\ \bibinfo {author} {\bibfnamefont {E.}~\bibnamefont {Kalnay}},\ }\href@noop {} {\bibfield  {journal} {\bibinfo  {journal} {PLOS ONE}\ }\textbf {\bibinfo {volume} {10}},\ \bibinfo {pages} {e0131226} (\bibinfo {year} {2015})}\BibitemShut {NoStop}%
\bibitem [{\citenamefont {Lagemann}\ \emph {et~al.}(2023)\citenamefont {Lagemann}, \citenamefont {Lagemann}, \citenamefont {Taschler},\ and\ \citenamefont {Mukherjee}}]{lagemann2023deep}%
  \BibitemOpen
  \bibfield  {author} {\bibinfo {author} {\bibfnamefont {K.}~\bibnamefont {Lagemann}}, \bibinfo {author} {\bibfnamefont {C.}~\bibnamefont {Lagemann}}, \bibinfo {author} {\bibfnamefont {B.}~\bibnamefont {Taschler}}, \ and\ \bibinfo {author} {\bibfnamefont {S.}~\bibnamefont {Mukherjee}},\ }\href@noop {} {\bibfield  {journal} {\bibinfo  {journal} {Nature Machine Intelligence}\ }\textbf {\bibinfo {volume} {5}},\ \bibinfo {pages} {1306} (\bibinfo {year} {2023})}\BibitemShut {NoStop}%
\bibitem [{\citenamefont {Wism{\"u}ller}\ \emph {et~al.}(2021)\citenamefont {Wism{\"u}ller}, \citenamefont {Dsouza}, \citenamefont {Vosoughi},\ and\ \citenamefont {Abidin}}]{wismuller2021large}%
  \BibitemOpen
  \bibfield  {author} {\bibinfo {author} {\bibfnamefont {A.}~\bibnamefont {Wism{\"u}ller}}, \bibinfo {author} {\bibfnamefont {A.~M.}\ \bibnamefont {Dsouza}}, \bibinfo {author} {\bibfnamefont {M.~A.}\ \bibnamefont {Vosoughi}}, \ and\ \bibinfo {author} {\bibfnamefont {A.}~\bibnamefont {Abidin}},\ }\href@noop {} {\bibfield  {journal} {\bibinfo  {journal} {Scientific reports}\ }\textbf {\bibinfo {volume} {11}},\ \bibinfo {pages} {7817} (\bibinfo {year} {2021})}\BibitemShut {NoStop}%
\bibitem [{\citenamefont {Keyes}\ \emph {et~al.}(2023)\citenamefont {Keyes}, \citenamefont {Giorgini},\ and\ \citenamefont {Wettlaufer}}]{keyes2023stochastic}%
  \BibitemOpen
  \bibfield  {author} {\bibinfo {author} {\bibfnamefont {N.~D.~B.}\ \bibnamefont {Keyes}}, \bibinfo {author} {\bibfnamefont {L.~T.}\ \bibnamefont {Giorgini}}, \ and\ \bibinfo {author} {\bibfnamefont {J.~S.}\ \bibnamefont {Wettlaufer}},\ }\href@noop {} {\bibfield  {journal} {\bibinfo  {journal} {Chaos: An Interdisciplinary Journal of Nonlinear Science}\ }\textbf {\bibinfo {volume} {33}},\ \bibinfo {pages} {093132} (\bibinfo {year} {2023})}\BibitemShut {NoStop}%
\bibitem [{\citenamefont {Aurell}\ and\ \citenamefont {Del~Ferraro}(2016)}]{aurell2016causal}%
  \BibitemOpen
  \bibfield  {author} {\bibinfo {author} {\bibfnamefont {E.}~\bibnamefont {Aurell}}\ and\ \bibinfo {author} {\bibfnamefont {G.}~\bibnamefont {Del~Ferraro}},\ }in\ \href@noop {} {\emph {\bibinfo {booktitle} {Journal of Physics: Conference Series}}},\ Vol.\ \bibinfo {volume} {699}\ (\bibinfo {organization} {IOP Publishing},\ \bibinfo {year} {2016})\ p.\ \bibinfo {pages} {012002}\BibitemShut {NoStop}%
\bibitem [{\citenamefont {Friedrich}\ \emph {et~al.}(2011)\citenamefont {Friedrich}, \citenamefont {Peinke}, \citenamefont {Sahimi},\ and\ \citenamefont {Tabar}}]{friedrich2011approaching}%
  \BibitemOpen
  \bibfield  {author} {\bibinfo {author} {\bibfnamefont {R.}~\bibnamefont {Friedrich}}, \bibinfo {author} {\bibfnamefont {J.}~\bibnamefont {Peinke}}, \bibinfo {author} {\bibfnamefont {M.}~\bibnamefont {Sahimi}}, \ and\ \bibinfo {author} {\bibfnamefont {M.~R.~R.}\ \bibnamefont {Tabar}},\ }\href@noop {} {\bibfield  {journal} {\bibinfo  {journal} {Physics Reports}\ }\textbf {\bibinfo {volume} {506}},\ \bibinfo {pages} {87} (\bibinfo {year} {2011})}\BibitemShut {NoStop}%
\bibitem [{\citenamefont {Granger}(1969)}]{granger1969investigating}%
  \BibitemOpen
  \bibfield  {author} {\bibinfo {author} {\bibfnamefont {C.~W.}\ \bibnamefont {Granger}},\ }\href@noop {} {\bibfield  {journal} {\bibinfo  {journal} {Econometrica: journal of the Econometric Society}\ ,\ \bibinfo {pages} {424}} (\bibinfo {year} {1969})}\BibitemShut {NoStop}%
\bibitem [{\citenamefont {Schreiber}(2000)}]{schreiber2000measuring}%
  \BibitemOpen
  \bibfield  {author} {\bibinfo {author} {\bibfnamefont {T.}~\bibnamefont {Schreiber}},\ }\href@noop {} {\bibfield  {journal} {\bibinfo  {journal} {Physical review letters}\ }\textbf {\bibinfo {volume} {85}},\ \bibinfo {pages} {461} (\bibinfo {year} {2000})}\BibitemShut {NoStop}%
\bibitem [{\citenamefont {Pearl}(2009)}]{pearl2009causal}%
  \BibitemOpen
  \bibfield  {author} {\bibinfo {author} {\bibfnamefont {J.}~\bibnamefont {Pearl}},\ }\href {\doibase 10.1214/09-SS057} {\bibfield  {journal} {\bibinfo  {journal} {Statistics Surveys}\ }\textbf {\bibinfo {volume} {3}},\ \bibinfo {pages} {96 } (\bibinfo {year} {2009})}\BibitemShut {NoStop}%
\bibitem [{\citenamefont {Camps-Valls}\ \emph {et~al.}(2023)\citenamefont {Camps-Valls}, \citenamefont {Gerhardus}, \citenamefont {Ninad}, \citenamefont {Varando}, \citenamefont {Martius}, \citenamefont {Balaguer-Ballester}, \citenamefont {Vinuesa}, \citenamefont {Diaz}, \citenamefont {Zanna},\ and\ \citenamefont {Runge}}]{camps2023discovering}%
  \BibitemOpen
  \bibfield  {author} {\bibinfo {author} {\bibfnamefont {G.}~\bibnamefont {Camps-Valls}}, \bibinfo {author} {\bibfnamefont {A.}~\bibnamefont {Gerhardus}}, \bibinfo {author} {\bibfnamefont {U.}~\bibnamefont {Ninad}}, \bibinfo {author} {\bibfnamefont {G.}~\bibnamefont {Varando}}, \bibinfo {author} {\bibfnamefont {G.}~\bibnamefont {Martius}}, \bibinfo {author} {\bibfnamefont {E.}~\bibnamefont {Balaguer-Ballester}}, \bibinfo {author} {\bibfnamefont {R.}~\bibnamefont {Vinuesa}}, \bibinfo {author} {\bibfnamefont {E.}~\bibnamefont {Diaz}}, \bibinfo {author} {\bibfnamefont {L.}~\bibnamefont {Zanna}}, \ and\ \bibinfo {author} {\bibfnamefont {J.}~\bibnamefont {Runge}},\ }\href@noop {} {\bibfield  {journal} {\bibinfo  {journal} {arXiv preprint arXiv:2305.13341}\ } (\bibinfo {year} {2023})}\BibitemShut {NoStop}%
\bibitem [{\citenamefont {Kaddour}\ \emph {et~al.}(2022)\citenamefont {Kaddour}, \citenamefont {Lynch}, \citenamefont {Liu}, \citenamefont {Kusner},\ and\ \citenamefont {Silva}}]{kaddour2022causal}%
  \BibitemOpen
  \bibfield  {author} {\bibinfo {author} {\bibfnamefont {J.}~\bibnamefont {Kaddour}}, \bibinfo {author} {\bibfnamefont {A.}~\bibnamefont {Lynch}}, \bibinfo {author} {\bibfnamefont {Q.}~\bibnamefont {Liu}}, \bibinfo {author} {\bibfnamefont {M.~J.}\ \bibnamefont {Kusner}}, \ and\ \bibinfo {author} {\bibfnamefont {R.}~\bibnamefont {Silva}},\ }\href@noop {} {\bibfield  {journal} {\bibinfo  {journal} {arXiv preprint arXiv:2206.15475}\ } (\bibinfo {year} {2022})}\BibitemShut {NoStop}%
\bibitem [{\citenamefont {Baldovin}\ \emph {et~al.}(2022)\citenamefont {Baldovin}, \citenamefont {Cecconi}, \citenamefont {Provenzale},\ and\ \citenamefont {Vulpiani}}]{baldovin2022extracting}%
  \BibitemOpen
  \bibfield  {author} {\bibinfo {author} {\bibfnamefont {M.}~\bibnamefont {Baldovin}}, \bibinfo {author} {\bibfnamefont {F.}~\bibnamefont {Cecconi}}, \bibinfo {author} {\bibfnamefont {A.}~\bibnamefont {Provenzale}}, \ and\ \bibinfo {author} {\bibfnamefont {A.}~\bibnamefont {Vulpiani}},\ }\href@noop {} {\bibfield  {journal} {\bibinfo  {journal} {Scientific Reports}\ }\textbf {\bibinfo {volume} {12}},\ \bibinfo {pages} {15320} (\bibinfo {year} {2022})}\BibitemShut {NoStop}%
\bibitem [{\citenamefont {Cecconi}\ \emph {et~al.}(2023)\citenamefont {Cecconi}, \citenamefont {Costantini}, \citenamefont {Guardiani}, \citenamefont {Baldovin},\ and\ \citenamefont {Vulpiani}}]{Cecconi_2023}%
  \BibitemOpen
  \bibfield  {author} {\bibinfo {author} {\bibfnamefont {F.}~\bibnamefont {Cecconi}}, \bibinfo {author} {\bibfnamefont {G.}~\bibnamefont {Costantini}}, \bibinfo {author} {\bibfnamefont {C.}~\bibnamefont {Guardiani}}, \bibinfo {author} {\bibfnamefont {M.}~\bibnamefont {Baldovin}}, \ and\ \bibinfo {author} {\bibfnamefont {A.}~\bibnamefont {Vulpiani}},\ }\href {\doibase 10.1088/1478-3975/ace1c5} {\bibfield  {journal} {\bibinfo  {journal} {Physical Biology}\ }\textbf {\bibinfo {volume} {20}},\ \bibinfo {pages} {056002} (\bibinfo {year} {2023})}\BibitemShut {NoStop}%
\bibitem [{\citenamefont {Falasca}\ \emph {et~al.}(2024)\citenamefont {Falasca}, \citenamefont {Perezhogin},\ and\ \citenamefont {Zanna}}]{falasca2023causal}%
  \BibitemOpen
  \bibfield  {author} {\bibinfo {author} {\bibfnamefont {F.}~\bibnamefont {Falasca}}, \bibinfo {author} {\bibfnamefont {P.}~\bibnamefont {Perezhogin}}, \ and\ \bibinfo {author} {\bibfnamefont {L.}~\bibnamefont {Zanna}},\ }\href {\doibase 10.1103/PhysRevE.109.044202} {\bibfield  {journal} {\bibinfo  {journal} {Phys. Rev. E}\ }\textbf {\bibinfo {volume} {109}},\ \bibinfo {pages} {044202} (\bibinfo {year} {2024})}\BibitemShut {NoStop}%
\bibitem [{\citenamefont {Baldovin}\ \emph {et~al.}(2020)\citenamefont {Baldovin}, \citenamefont {Cecconi},\ and\ \citenamefont {Vulpiani}}]{baldovin2020understanding}%
  \BibitemOpen
  \bibfield  {author} {\bibinfo {author} {\bibfnamefont {M.}~\bibnamefont {Baldovin}}, \bibinfo {author} {\bibfnamefont {F.}~\bibnamefont {Cecconi}}, \ and\ \bibinfo {author} {\bibfnamefont {A.}~\bibnamefont {Vulpiani}},\ }\href@noop {} {\bibfield  {journal} {\bibinfo  {journal} {Physical Review Research}\ }\textbf {\bibinfo {volume} {2}},\ \bibinfo {pages} {043436} (\bibinfo {year} {2020})}\BibitemShut {NoStop}%
\bibitem [{\citenamefont {Lucarini}(2018)}]{lucarini2018revising}%
  \BibitemOpen
  \bibfield  {author} {\bibinfo {author} {\bibfnamefont {V.}~\bibnamefont {Lucarini}},\ }\href@noop {} {\bibfield  {journal} {\bibinfo  {journal} {Journal of Statistical Physics}\ }\textbf {\bibinfo {volume} {173}},\ \bibinfo {pages} {1698} (\bibinfo {year} {2018})}\BibitemShut {NoStop}%
\bibitem [{\citenamefont {Marconi}\ \emph {et~al.}(2008)\citenamefont {Marconi}, \citenamefont {Puglisi}, \citenamefont {Rondoni},\ and\ \citenamefont {Vulpiani}}]{marconi2008fluctuation}%
  \BibitemOpen
  \bibfield  {author} {\bibinfo {author} {\bibfnamefont {U.~M.~B.}\ \bibnamefont {Marconi}}, \bibinfo {author} {\bibfnamefont {A.}~\bibnamefont {Puglisi}}, \bibinfo {author} {\bibfnamefont {L.}~\bibnamefont {Rondoni}}, \ and\ \bibinfo {author} {\bibfnamefont {A.}~\bibnamefont {Vulpiani}},\ }\href@noop {} {\bibfield  {journal} {\bibinfo  {journal} {Physics reports}\ }\textbf {\bibinfo {volume} {461}},\ \bibinfo {pages} {111} (\bibinfo {year} {2008})}\BibitemShut {NoStop}%
\bibitem [{\citenamefont {Daum}(2005)}]{daum2005nonlinear}%
  \BibitemOpen
  \bibfield  {author} {\bibinfo {author} {\bibfnamefont {F.}~\bibnamefont {Daum}},\ }\href@noop {} {\bibfield  {journal} {\bibinfo  {journal} {IEEE Aerospace and Electronic Systems Magazine}\ }\textbf {\bibinfo {volume} {20}},\ \bibinfo {pages} {57} (\bibinfo {year} {2005})}\BibitemShut {NoStop}%
\bibitem [{\citenamefont {Sj{\"o}berg}\ \emph {et~al.}(2009)\citenamefont {Sj{\"o}berg}, \citenamefont {L{\"o}tstedt},\ and\ \citenamefont {Elf}}]{sjoberg2009fokker}%
  \BibitemOpen
  \bibfield  {author} {\bibinfo {author} {\bibfnamefont {P.}~\bibnamefont {Sj{\"o}berg}}, \bibinfo {author} {\bibfnamefont {P.}~\bibnamefont {L{\"o}tstedt}}, \ and\ \bibinfo {author} {\bibfnamefont {J.}~\bibnamefont {Elf}},\ }\href@noop {} {\bibfield  {journal} {\bibinfo  {journal} {Computing and Visualization in Science}\ }\textbf {\bibinfo {volume} {12}},\ \bibinfo {pages} {37} (\bibinfo {year} {2009})}\BibitemShut {NoStop}%
\bibitem [{\citenamefont {Cooper}\ and\ \citenamefont {Haynes}(2011)}]{ClimateSensitivityviaaNonparametricFluctuationDissipationTheorem}%
  \BibitemOpen
  \bibfield  {author} {\bibinfo {author} {\bibfnamefont {F.~C.}\ \bibnamefont {Cooper}}\ and\ \bibinfo {author} {\bibfnamefont {P.~H.}\ \bibnamefont {Haynes}},\ }\href {\doibase 10.1175/2010JAS3633.1} {\bibfield  {journal} {\bibinfo  {journal} {Journal of the Atmospheric Sciences}\ }\textbf {\bibinfo {volume} {68}},\ \bibinfo {pages} {937 } (\bibinfo {year} {2011})}\BibitemShut {NoStop}%
\bibitem [{\citenamefont {Leith}(1975)}]{ClimateResponseandFluctuationDissipation}%
  \BibitemOpen
  \bibfield  {author} {\bibinfo {author} {\bibfnamefont {C.~E.}\ \bibnamefont {Leith}},\ }\href {\doibase 10.1175/1520-0469(1975)032<2022:CRAFD>2.0.CO;2} {\bibfield  {journal} {\bibinfo  {journal} {Journal of Atmospheric Sciences}\ }\textbf {\bibinfo {volume} {32}},\ \bibinfo {pages} {2022 } (\bibinfo {year} {1975})}\BibitemShut {NoStop}%
\bibitem [{\citenamefont {Sardeshmukh}\ and\ \citenamefont {Sura}(2009)}]{ReconcilingNonGaussianClimateStatisticswithLinearDynamics}%
  \BibitemOpen
  \bibfield  {author} {\bibinfo {author} {\bibfnamefont {P.~D.}\ \bibnamefont {Sardeshmukh}}\ and\ \bibinfo {author} {\bibfnamefont {P.}~\bibnamefont {Sura}},\ }\href {\doibase 10.1175/2008JCLI2358.1} {\bibfield  {journal} {\bibinfo  {journal} {Journal of Climate}\ }\textbf {\bibinfo {volume} {22}},\ \bibinfo {pages} {1193 } (\bibinfo {year} {2009})}\BibitemShut {NoStop}%
\bibitem [{\citenamefont {Gritsun}\ and\ \citenamefont {Lucarini}(2017)}]{GRITSUN201762}%
  \BibitemOpen
  \bibfield  {author} {\bibinfo {author} {\bibfnamefont {A.}~\bibnamefont {Gritsun}}\ and\ \bibinfo {author} {\bibfnamefont {V.}~\bibnamefont {Lucarini}},\ }\href {\doibase https://doi.org/10.1016/j.physd.2017.02.015} {\bibfield  {journal} {\bibinfo  {journal} {Physica D: Nonlinear Phenomena}\ }\textbf {\bibinfo {volume} {349}},\ \bibinfo {pages} {62} (\bibinfo {year} {2017})}\BibitemShut {NoStop}%
\bibitem [{\citenamefont {Proistosescu}\ \emph {et~al.}(2016)\citenamefont {Proistosescu}, \citenamefont {Rhines},\ and\ \citenamefont {Huybers}}]{proistosescu2016identification}%
  \BibitemOpen
  \bibfield  {author} {\bibinfo {author} {\bibfnamefont {C.}~\bibnamefont {Proistosescu}}, \bibinfo {author} {\bibfnamefont {A.}~\bibnamefont {Rhines}}, \ and\ \bibinfo {author} {\bibfnamefont {P.}~\bibnamefont {Huybers}},\ }\href@noop {} {\bibfield  {journal} {\bibinfo  {journal} {Geophysical Research Letters}\ }\textbf {\bibinfo {volume} {43}},\ \bibinfo {pages} {5425} (\bibinfo {year} {2016})}\BibitemShut {NoStop}%
\bibitem [{\citenamefont {Loikith}\ and\ \citenamefont {Neelin}(2015)}]{loikith2015short}%
  \BibitemOpen
  \bibfield  {author} {\bibinfo {author} {\bibfnamefont {P.~C.}\ \bibnamefont {Loikith}}\ and\ \bibinfo {author} {\bibfnamefont {J.~D.}\ \bibnamefont {Neelin}},\ }\href@noop {} {\bibfield  {journal} {\bibinfo  {journal} {Geophysical Research Letters}\ }\textbf {\bibinfo {volume} {42}},\ \bibinfo {pages} {8577} (\bibinfo {year} {2015})}\BibitemShut {NoStop}%
\bibitem [{\citenamefont {Branicki}\ and\ \citenamefont {Majda}(2012)}]{branicki2012quantifying}%
  \BibitemOpen
  \bibfield  {author} {\bibinfo {author} {\bibfnamefont {M.}~\bibnamefont {Branicki}}\ and\ \bibinfo {author} {\bibfnamefont {A.~J.}\ \bibnamefont {Majda}},\ }\href@noop {} {\bibfield  {journal} {\bibinfo  {journal} {Nonlinearity}\ }\textbf {\bibinfo {volume} {25}},\ \bibinfo {pages} {2543} (\bibinfo {year} {2012})}\BibitemShut {NoStop}%
\bibitem [{\citenamefont {Martinez-Villalobos}\ and\ \citenamefont {Neelin}(2019)}]{WhyDoPrecipitationIntensitiesTendtoFollowGammaDistributions}%
  \BibitemOpen
  \bibfield  {author} {\bibinfo {author} {\bibfnamefont {C.}~\bibnamefont {Martinez-Villalobos}}\ and\ \bibinfo {author} {\bibfnamefont {J.~D.}\ \bibnamefont {Neelin}},\ }\href {\doibase 10.1175/JAS-D-18-0343.1} {\bibfield  {journal} {\bibinfo  {journal} {Journal of the Atmospheric Sciences}\ }\textbf {\bibinfo {volume} {76}},\ \bibinfo {pages} {3611 } (\bibinfo {year} {2019})}\BibitemShut {NoStop}%
\bibitem [{\citenamefont {Kriegeskorte}\ and\ \citenamefont {Wei}(2021)}]{kriegeskorte2021neural}%
  \BibitemOpen
  \bibfield  {author} {\bibinfo {author} {\bibfnamefont {N.}~\bibnamefont {Kriegeskorte}}\ and\ \bibinfo {author} {\bibfnamefont {X.-X.}\ \bibnamefont {Wei}},\ }\href@noop {} {\bibfield  {journal} {\bibinfo  {journal} {Nature Reviews Neuroscience}\ }\textbf {\bibinfo {volume} {22}},\ \bibinfo {pages} {703} (\bibinfo {year} {2021})}\BibitemShut {NoStop}%
\bibitem [{\citenamefont {Vincent}(2011)}]{Vincent2011}%
  \BibitemOpen
  \bibfield  {author} {\bibinfo {author} {\bibfnamefont {P.}~\bibnamefont {Vincent}},\ }\href@noop {} {\bibfield  {journal} {\bibinfo  {journal} {Neural computation}\ }\textbf {\bibinfo {volume} {23}},\ \bibinfo {pages} {1661} (\bibinfo {year} {2011})}\BibitemShut {NoStop}%
\bibitem [{\citenamefont {Ho}\ \emph {et~al.}(2020)\citenamefont {Ho}, \citenamefont {Jain},\ and\ \citenamefont {Abbeel}}]{Ho2020}%
  \BibitemOpen
  \bibfield  {author} {\bibinfo {author} {\bibfnamefont {J.}~\bibnamefont {Ho}}, \bibinfo {author} {\bibfnamefont {A.}~\bibnamefont {Jain}}, \ and\ \bibinfo {author} {\bibfnamefont {P.}~\bibnamefont {Abbeel}},\ }\href@noop {} {\bibfield  {journal} {\bibinfo  {journal} {Advances in neural information processing systems}\ }\textbf {\bibinfo {volume} {33}},\ \bibinfo {pages} {6840} (\bibinfo {year} {2020})}\BibitemShut {NoStop}%
\bibitem [{\citenamefont {Song}\ and\ \citenamefont {Ermon}(2019)}]{Song2019}%
  \BibitemOpen
  \bibfield  {author} {\bibinfo {author} {\bibfnamefont {Y.}~\bibnamefont {Song}}\ and\ \bibinfo {author} {\bibfnamefont {S.}~\bibnamefont {Ermon}},\ }\href@noop {} {\bibfield  {journal} {\bibinfo  {journal} {Advances in neural information processing systems}\ }\textbf {\bibinfo {volume} {32}} (\bibinfo {year} {2019})}\BibitemShut {NoStop}%
\bibitem [{\citenamefont {Song}\ \emph {et~al.}(2020)\citenamefont {Song}, \citenamefont {Sohl-Dickstein}, \citenamefont {Kingma}, \citenamefont {Kumar}, \citenamefont {Ermon},\ and\ \citenamefont {Poole}}]{song2020score}%
  \BibitemOpen
  \bibfield  {author} {\bibinfo {author} {\bibfnamefont {Y.}~\bibnamefont {Song}}, \bibinfo {author} {\bibfnamefont {J.}~\bibnamefont {Sohl-Dickstein}}, \bibinfo {author} {\bibfnamefont {D.~P.}\ \bibnamefont {Kingma}}, \bibinfo {author} {\bibfnamefont {A.}~\bibnamefont {Kumar}}, \bibinfo {author} {\bibfnamefont {S.}~\bibnamefont {Ermon}}, \ and\ \bibinfo {author} {\bibfnamefont {B.}~\bibnamefont {Poole}},\ }\href@noop {} {\bibfield  {journal} {\bibinfo  {journal} {arXiv preprint arXiv:2011.13456}\ } (\bibinfo {year} {2020})}\BibitemShut {NoStop}%
\bibitem [{\citenamefont {Bischoff}\ and\ \citenamefont {Deck}(2023)}]{Bischoff2023}%
  \BibitemOpen
  \bibfield  {author} {\bibinfo {author} {\bibfnamefont {T.}~\bibnamefont {Bischoff}}\ and\ \bibinfo {author} {\bibfnamefont {K.}~\bibnamefont {Deck}},\ }\href@noop {} {\bibfield  {journal} {\bibinfo  {journal} {arXiv preprint arXiv:2305.01822}\ } (\bibinfo {year} {2023})}\BibitemShut {NoStop}%
\bibitem [{\citenamefont {Liu}\ \emph {et~al.}(2023)\citenamefont {Liu}, \citenamefont {Luo}, \citenamefont {Xu}, \citenamefont {Jaakkola},\ and\ \citenamefont {Tegmark}}]{liu2023genphys}%
  \BibitemOpen
  \bibfield  {author} {\bibinfo {author} {\bibfnamefont {Z.}~\bibnamefont {Liu}}, \bibinfo {author} {\bibfnamefont {D.}~\bibnamefont {Luo}}, \bibinfo {author} {\bibfnamefont {Y.}~\bibnamefont {Xu}}, \bibinfo {author} {\bibfnamefont {T.}~\bibnamefont {Jaakkola}}, \ and\ \bibinfo {author} {\bibfnamefont {M.}~\bibnamefont {Tegmark}},\ }\href@noop {} {\bibfield  {journal} {\bibinfo  {journal} {arXiv preprint arXiv:2304.02637}\ } (\bibinfo {year} {2023})}\BibitemShut {NoStop}%
\bibitem [{\citenamefont {Stanczuk}\ \emph {et~al.}(2023)\citenamefont {Stanczuk}, \citenamefont {Batzolis}, \citenamefont {Deveney},\ and\ \citenamefont {Schönlieb}}]{stanczuk2023diffusion}%
  \BibitemOpen
  \bibfield  {author} {\bibinfo {author} {\bibfnamefont {J.}~\bibnamefont {Stanczuk}}, \bibinfo {author} {\bibfnamefont {G.}~\bibnamefont {Batzolis}}, \bibinfo {author} {\bibfnamefont {T.}~\bibnamefont {Deveney}}, \ and\ \bibinfo {author} {\bibfnamefont {C.-B.}\ \bibnamefont {Schönlieb}},\ }\href@noop {} {\enquote {\bibinfo {title} {Your diffusion model secretly knows the dimension of the data manifold},}\ } (\bibinfo {year} {2023}),\ \Eprint {http://arxiv.org/abs/2212.12611} {arXiv:2212.12611 [cs.LG]} \BibitemShut {NoStop}%
\bibitem [{\citenamefont {Ronneberger}\ \emph {et~al.}(2015)\citenamefont {Ronneberger}, \citenamefont {Fischer},\ and\ \citenamefont {Brox}}]{Ronneberger2015}%
  \BibitemOpen
  \bibfield  {author} {\bibinfo {author} {\bibfnamefont {O.}~\bibnamefont {Ronneberger}}, \bibinfo {author} {\bibfnamefont {P.}~\bibnamefont {Fischer}}, \ and\ \bibinfo {author} {\bibfnamefont {T.}~\bibnamefont {Brox}},\ }in\ \href@noop {} {\emph {\bibinfo {booktitle} {Medical Image Computing and Computer-Assisted Intervention--MICCAI 2015: 18th International Conference, Munich, Germany, October 5-9, 2015, Proceedings, Part III 18}}}\ (\bibinfo {organization} {Springer},\ \bibinfo {year} {2015})\ pp.\ \bibinfo {pages} {234--241}\BibitemShut {NoStop}%
\bibitem [{\citenamefont {Adcock}\ \emph {et~al.}(2020)\citenamefont {Adcock}, \citenamefont {Brugiapaglia}, \citenamefont {Dexter},\ and\ \citenamefont {Moraga}}]{adcock2020deep}%
  \BibitemOpen
  \bibfield  {author} {\bibinfo {author} {\bibfnamefont {B.}~\bibnamefont {Adcock}}, \bibinfo {author} {\bibfnamefont {S.}~\bibnamefont {Brugiapaglia}}, \bibinfo {author} {\bibfnamefont {N.}~\bibnamefont {Dexter}}, \ and\ \bibinfo {author} {\bibfnamefont {S.}~\bibnamefont {Moraga}},\ }\href@noop {} {\bibfield  {journal} {\bibinfo  {journal} {arXiv preprint arXiv:2012.06081}\ } (\bibinfo {year} {2020})}\BibitemShut {NoStop}%
\bibitem [{\citenamefont {Beneventano}\ \emph {et~al.}(2021)\citenamefont {Beneventano}, \citenamefont {Cheridito}, \citenamefont {Graeber}, \citenamefont {Jentzen},\ and\ \citenamefont {Kuckuck}}]{beneventano2021deep}%
  \BibitemOpen
  \bibfield  {author} {\bibinfo {author} {\bibfnamefont {P.}~\bibnamefont {Beneventano}}, \bibinfo {author} {\bibfnamefont {P.}~\bibnamefont {Cheridito}}, \bibinfo {author} {\bibfnamefont {R.}~\bibnamefont {Graeber}}, \bibinfo {author} {\bibfnamefont {A.}~\bibnamefont {Jentzen}}, \ and\ \bibinfo {author} {\bibfnamefont {B.}~\bibnamefont {Kuckuck}},\ }\href@noop {} {\bibfield  {journal} {\bibinfo  {journal} {arXiv preprint arXiv:2112.14523}\ } (\bibinfo {year} {2021})}\BibitemShut {NoStop}%
\bibitem [{\citenamefont {Gritsun}\ and\ \citenamefont {Branstator}(2007)}]{gritsun2007climate}%
  \BibitemOpen
  \bibfield  {author} {\bibinfo {author} {\bibfnamefont {A.}~\bibnamefont {Gritsun}}\ and\ \bibinfo {author} {\bibfnamefont {G.}~\bibnamefont {Branstator}},\ }\href@noop {} {\bibfield  {journal} {\bibinfo  {journal} {Journal of the atmospheric sciences}\ }\textbf {\bibinfo {volume} {64}},\ \bibinfo {pages} {2558} (\bibinfo {year} {2007})}\BibitemShut {NoStop}%
\bibitem [{\citenamefont {Uhlenbeck}\ and\ \citenamefont {Ornstein}(1930)}]{uhlenbeck1930theory}%
  \BibitemOpen
  \bibfield  {author} {\bibinfo {author} {\bibfnamefont {G.~E.}\ \bibnamefont {Uhlenbeck}}\ and\ \bibinfo {author} {\bibfnamefont {L.~S.}\ \bibnamefont {Ornstein}},\ }\href@noop {} {\bibfield  {journal} {\bibinfo  {journal} {Physical review}\ }\textbf {\bibinfo {volume} {36}},\ \bibinfo {pages} {823} (\bibinfo {year} {1930})}\BibitemShut {NoStop}%
\bibitem [{\citenamefont {Maller}\ \emph {et~al.}(2009)\citenamefont {Maller}, \citenamefont {M{\"u}ller},\ and\ \citenamefont {Szimayer}}]{maller2009ornstein}%
  \BibitemOpen
  \bibfield  {author} {\bibinfo {author} {\bibfnamefont {R.~A.}\ \bibnamefont {Maller}}, \bibinfo {author} {\bibfnamefont {G.}~\bibnamefont {M{\"u}ller}}, \ and\ \bibinfo {author} {\bibfnamefont {A.}~\bibnamefont {Szimayer}},\ }\href@noop {} {\bibfield  {journal} {\bibinfo  {journal} {Handbook of financial time series}\ ,\ \bibinfo {pages} {421}} (\bibinfo {year} {2009})}\BibitemShut {NoStop}%
\bibitem [{\citenamefont {Griffin}\ and\ \citenamefont {Steel}(2006)}]{griffin2006inference}%
  \BibitemOpen
  \bibfield  {author} {\bibinfo {author} {\bibfnamefont {J.~E.}\ \bibnamefont {Griffin}}\ and\ \bibinfo {author} {\bibfnamefont {M.~F.}\ \bibnamefont {Steel}},\ }\href@noop {} {\bibfield  {journal} {\bibinfo  {journal} {Journal of Econometrics}\ }\textbf {\bibinfo {volume} {134}},\ \bibinfo {pages} {605} (\bibinfo {year} {2006})}\BibitemShut {NoStop}%
\bibitem [{\citenamefont {Frank}\ \emph {et~al.}(2000)\citenamefont {Frank}, \citenamefont {Daffertshofer},\ and\ \citenamefont {Beek}}]{frank2000multivariate}%
  \BibitemOpen
  \bibfield  {author} {\bibinfo {author} {\bibfnamefont {T.}~\bibnamefont {Frank}}, \bibinfo {author} {\bibfnamefont {A.}~\bibnamefont {Daffertshofer}}, \ and\ \bibinfo {author} {\bibfnamefont {P.}~\bibnamefont {Beek}},\ }\href@noop {} {\bibfield  {journal} {\bibinfo  {journal} {Physical Review E}\ }\textbf {\bibinfo {volume} {63}},\ \bibinfo {pages} {011905} (\bibinfo {year} {2000})}\BibitemShut {NoStop}%
\bibitem [{\citenamefont {Vatiwutipong}\ and\ \citenamefont {Phewchean}(2019)}]{vatiwutipong2019alternative}%
  \BibitemOpen
  \bibfield  {author} {\bibinfo {author} {\bibfnamefont {P.}~\bibnamefont {Vatiwutipong}}\ and\ \bibinfo {author} {\bibfnamefont {N.}~\bibnamefont {Phewchean}},\ }\href@noop {} {\bibfield  {journal} {\bibinfo  {journal} {Advances in Difference Equations}\ }\textbf {\bibinfo {volume} {2019}},\ \bibinfo {pages} {1} (\bibinfo {year} {2019})}\BibitemShut {NoStop}%
\bibitem [{\citenamefont {Allen}\ and\ \citenamefont {Cahn}(1972)}]{allen1972ground}%
  \BibitemOpen
  \bibfield  {author} {\bibinfo {author} {\bibfnamefont {S.~M.}\ \bibnamefont {Allen}}\ and\ \bibinfo {author} {\bibfnamefont {J.~W.}\ \bibnamefont {Cahn}},\ }\href@noop {} {\bibfield  {journal} {\bibinfo  {journal} {Acta Metallurgica}\ }\textbf {\bibinfo {volume} {20}},\ \bibinfo {pages} {423} (\bibinfo {year} {1972})}\BibitemShut {NoStop}%
\bibitem [{\citenamefont {Kolmogorov}\ \emph {et~al.}(1937)\citenamefont {Kolmogorov}, \citenamefont {Petrovskii},\ and\ \citenamefont {Piskunov}}]{kolmogorov1study}%
  \BibitemOpen
  \bibfield  {author} {\bibinfo {author} {\bibfnamefont {A.}~\bibnamefont {Kolmogorov}}, \bibinfo {author} {\bibfnamefont {I.}~\bibnamefont {Petrovskii}}, \ and\ \bibinfo {author} {\bibfnamefont {N.}~\bibnamefont {Piskunov}},\ }\href@noop {} {\bibfield  {journal} {\bibinfo  {journal} {Selected Works of AN Kolmogorov}\ }\textbf {\bibinfo {volume} {1}} (\bibinfo {year} {1937})}\BibitemShut {NoStop}%
\bibitem [{\citenamefont {Turing}(1990)}]{turing1990chemical}%
  \BibitemOpen
  \bibfield  {author} {\bibinfo {author} {\bibfnamefont {A.~M.}\ \bibnamefont {Turing}},\ }\href@noop {} {\bibfield  {journal} {\bibinfo  {journal} {Bulletin of mathematical biology}\ }\textbf {\bibinfo {volume} {52}},\ \bibinfo {pages} {153} (\bibinfo {year} {1990})}\BibitemShut {NoStop}%
\bibitem [{\citenamefont {Callahan}\ and\ \citenamefont {Knobloch}(1999)}]{callahan1999pattern}%
  \BibitemOpen
  \bibfield  {author} {\bibinfo {author} {\bibfnamefont {T.}~\bibnamefont {Callahan}}\ and\ \bibinfo {author} {\bibfnamefont {E.}~\bibnamefont {Knobloch}},\ }\href@noop {} {\bibfield  {journal} {\bibinfo  {journal} {Physica D: Nonlinear Phenomena}\ }\textbf {\bibinfo {volume} {132}},\ \bibinfo {pages} {339} (\bibinfo {year} {1999})}\BibitemShut {NoStop}%
\bibitem [{\citenamefont {Kondo}\ and\ \citenamefont {Miura}(2010)}]{kondo2010reaction}%
  \BibitemOpen
  \bibfield  {author} {\bibinfo {author} {\bibfnamefont {S.}~\bibnamefont {Kondo}}\ and\ \bibinfo {author} {\bibfnamefont {T.}~\bibnamefont {Miura}},\ }\href@noop {} {\bibfield  {journal} {\bibinfo  {journal} {science}\ }\textbf {\bibinfo {volume} {329}},\ \bibinfo {pages} {1616} (\bibinfo {year} {2010})}\BibitemShut {NoStop}%
\bibitem [{\citenamefont {Flierl}\ and\ \citenamefont {Souza}(2024)}]{Flierl_Souza_2024}%
  \BibitemOpen
  \bibfield  {author} {\bibinfo {author} {\bibfnamefont {G.~R.}\ \bibnamefont {Flierl}}\ and\ \bibinfo {author} {\bibfnamefont {A.~N.}\ \bibnamefont {Souza}},\ }\href {\doibase 10.1017/jfm.2024.302} {\bibfield  {journal} {\bibinfo  {journal} {Journal of Fluid Mechanics}\ }\textbf {\bibinfo {volume} {986}},\ \bibinfo {pages} {A8} (\bibinfo {year} {2024})}\BibitemShut {NoStop}%
\bibitem [{\citenamefont {Majda}\ \emph {et~al.}(2010)\citenamefont {Majda}, \citenamefont {Gershgorin},\ and\ \citenamefont {Yuan}}]{LowFrequencyClimateResponseandFluctuationDissipationTheoremsTheoryandPractice}%
  \BibitemOpen
  \bibfield  {author} {\bibinfo {author} {\bibfnamefont {A.~J.}\ \bibnamefont {Majda}}, \bibinfo {author} {\bibfnamefont {B.}~\bibnamefont {Gershgorin}}, \ and\ \bibinfo {author} {\bibfnamefont {Y.}~\bibnamefont {Yuan}},\ }\href {\doibase 10.1175/2009JAS3264.1} {\bibfield  {journal} {\bibinfo  {journal} {Journal of the Atmospheric Sciences}\ }\textbf {\bibinfo {volume} {67}},\ \bibinfo {pages} {1186 } (\bibinfo {year} {2010})}\BibitemShut {NoStop}%
\bibitem [{\citenamefont {Gershgorin}\ and\ \citenamefont {Majda}(2010)}]{GERSHGORIN20101741}%
  \BibitemOpen
  \bibfield  {author} {\bibinfo {author} {\bibfnamefont {B.}~\bibnamefont {Gershgorin}}\ and\ \bibinfo {author} {\bibfnamefont {A.~J.}\ \bibnamefont {Majda}},\ }\href {\doibase https://doi.org/10.1016/j.physd.2010.05.009} {\bibfield  {journal} {\bibinfo  {journal} {Physica D: Nonlinear Phenomena}\ }\textbf {\bibinfo {volume} {239}},\ \bibinfo {pages} {1741} (\bibinfo {year} {2010})}\BibitemShut {NoStop}%
\bibitem [{\citenamefont {Majda}\ and\ \citenamefont {Qi}(2019)}]{Majda2019}%
  \BibitemOpen
  \bibfield  {author} {\bibinfo {author} {\bibfnamefont {A.~J.}\ \bibnamefont {Majda}}\ and\ \bibinfo {author} {\bibfnamefont {D.}~\bibnamefont {Qi}},\ }\href {\doibase 10.1063/1.5118690} {\bibfield  {journal} {\bibinfo  {journal} {Chaos: An Interdisciplinary Journal of Nonlinear Science}\ }\textbf {\bibinfo {volume} {29}},\ \bibinfo {pages} {103131} (\bibinfo {year} {2019})},\ \Eprint {http://arxiv.org/abs/https://pubs.aip.org/aip/cha/article-pdf/doi/10.1063/1.5118690/14624899/103131\_1\_online.pdf} {https://pubs.aip.org/aip/cha/article-pdf/doi/10.1063/1.5118690/14624899/103131\_1\_online.pdf} \BibitemShut {NoStop}%
\end{thebibliography}%
\end{document}